\documentclass[twocolumn,preprint,final,3p,times]{elsarticle}
\usepackage{amsfonts}
\usepackage{amssymb}
\usepackage{amsmath}
\usepackage{graphicx}
\usepackage{bm}
\usepackage{color}
\usepackage{verbatim}
\usepackage{subfigure}

\begin{document}
\title{Active Matter in Lateral Parabolic Confinement: From Subdiffusion to Superdiffusion\tnoteref{t1,t2}}
\author[rvt]{H.E. Ribeiro}
\ead{he.ribeiro@bol.com.br}
%\email{fqpotiguar@ufpa.br}
\address[rvt]{Universidade Federal do Par\'a, Faculdade de F\'\i sica, ICEN, 
	Avenida Augusto Corr\^ea 1, Guam\'a, 66075-110, Bel\'em, Par\'a, Brazil}
\author[rvt]{F. Q. Potiguar}%\corref{cor2}}
%\ead{fqpotiguar@gmail.com}
%\address[focal]{River Valley Technologies, 9, Browns Court,
%	Kennford, Exeter, United Kingdom}

\begin{abstract}
	
In this work we studied the diffusive behavior of active brownian particles under lateral parabolic confinement. The results showed that we go from subdiffusion to ballistic motion as we vary the angular noise strength and confinement intensity. We argued that the subdiffusion regimes appear as consequence of the restricted space available for diffusion (achieved either through large confinement and/or large noise); we saw that when there are large confinement and noise intensity, a similar configuration to single file diffusion appears; on the other hand, normal and superdiffusive regimes may occur due to low noise (longer persistent motion), either through exploring a wider region around the potential minimum in the transverse direction (low confinement), or by forming independent clusters (high confinement).
\end{abstract}

\begin{keyword}
Active brownian particles; Angular noise; Confinement; Diffusion.	
\end{keyword}	
%\PACS{87.80.Fe, 47.63.Gd, 87.15.hj, 05.40.-a}

%keywords: Active brownian particles; Angular noise; Confinement; Normal diffusion; Superdiffusion; Subdiffusion.
\maketitle
% ======================================== INTRODUCTION  =====================================================
\section{Introduction}
Transport phenomena are well known in the physical world. Diffusion has received special attention by the scientific community. More specifically, diffusion of particles with presence of constraints has been intensely studied \cite{hahn}. In this respect, several works showed that particles moving under some kind of spatial restriction may exhibit anomalous diffusion \cite{lucena}; depending on the nature of the  confinement, it can happen that mutual passage among particles is forbidden, resulting in a phenomenon called single file diffusion (SFD). Some realizations of SFD include transport of water and ions through molecular-sized channels in biological membranes \cite{wei}, diffusion of colloids in microfluidic devices \cite{wei, meron}, molecular transport in zeolites \cite{hahn,gupta,sholl} and passage of molecules through narrow pores \cite{A. Hodgkin}, being this latter the first approach about SFD mechanism. Systems of particles limited to move in confined environments (or submitted to an adjoining external field) display a different dynamics from one expected for free particles \cite{Peeters}. In addition to confinement, the interparticle interaction also plays a relevant role in the description of this diffusion process \cite{Wen}, with the possibility of letting the diffusion even slower than in the SFD case \cite{Nelissen}.\\

The most common way, which by no means cover all cases, of characterizing a diffusion process is through the long-time limit of the mean square displacement (MSD), $\langle(\Delta x)^2\rangle$, which is: 
\begin{equation}
\langle(\Delta x)^2\rangle=Dt^\alpha \label{eq1},	
\end{equation}	                                               
with $D$ being the effective diffusion coefficient and $\alpha$ the diffusive exponent, which allows us to sort the diffusion process as normal ($\alpha=1$), superdiffusion ($\alpha>1$) or subdiffusion ($\alpha<1$); if $\alpha=2$, we have ballistic diffusion, and if $\alpha=0.5$, we have a characteristic behavior of SFD (however not deterministic). Some theoretical works have related crossover occurrences from normal to subdiffusion (including SFD \cite{D.lucena}) and between normal diffusion and ballistic one \cite{Francisco J. Sevilla}.\\

Lately, a novel type of model particles, called self-propelled or active matter \cite{Tamas Vicsek, potiguar, Fily} have been proposed, which are so-called due to the fact that they propel themselves through some internal mechanism. Recently, there has been an increasing interest in the transport properties of self-propelled particles \cite{Romanczuk,bechinger16}, more specifically, with diffusion in confined geometries \cite{burada}. Locatelli {\em et al}. \cite{locatelli} have studied the confinement effect in a pure one-dimensional active particle system, and the consequent change in the system dynamics, specially in the diffusion regimes \cite{lutz}. In this letter, we study the diffusion regimes, in the $x$ direction, of a system of active particles, confined by a parabolic potential in the perpendicular, i. .e $y$, direction, as a function of the confinement strength and rotational noise intensity. We are mainly interested in measuring the mean-squared displacement (MSD) of the active particles,
\begin{equation}
\langle\Delta x^2(\Delta t)\rangle=\left\langle\frac{1}{N}\sum_{i=1}^N[x_i(t+\Delta t)-x_i(t)]^2\right\rangle, \label{eq2}
\end{equation}   
in the perpendicular direction, and obtaining the diffusion, given by $\alpha$ in (\ref{eq1}). The results show that increasing the confinement and angular noise strengths, we go from superdiffusion (even ballistic in some cases) to subdiffusion (practically attaining the SFD regime).\\

% As a result, the system assumes different arrangements, more disperse or more concentrated around the potential minimum. given by probability density function (PDF). Also an important property for fluids is computed in order to confirming our forecasts about the tendency of clustering at low $D_r$ and thus to explaining the several curves for the MSD.\\

The paper is organized as follows. In Sec. \ref{Model}, we present the model employed in the simulations as well as the method used for integrating the equations of motion. In Sec. \ref{Results}, we show our main results for the MSD curves along with additional data to support our observations. Lastly, in Sec. \ref{Conclusions}, we present our conclusions.

% ======================================== MODEL =====================================================
%\noindent {\it Model}
\section{Model}\label{Model}
In our simulations, we employed the angular brownian motion (ABM) \cite{potiguar, Fily} model in a two-dimensional system of $N=1000$ particles. Initially they are randomly distributed in the simulation box, and have random velocity directions. The dynamics of the {\it i}-th particle is given by the Langevin equation in the limit of low Reynolds number (overdamped regime)     
\begin{equation}
 \label{eq_1}
\frac {\partial{\bf r}_i}{\partial t}={\textbf{\textit{v}}}_{i}+\mu{\bf F}_i+{\boldsymbol{\xi}}_i(t),$$$$
\frac{\partial \theta_i(t)}{\partial t}=\eta_i(t).\label{eq3}
 \end{equation}
 These equations are integrated via the stochastic second-order Runge-Kutta method (RKSII) \cite{rebecca},
 in which ${\bf r}_i$ is the vector position of each particle, $\theta_i(t)$ sets the direction of the vector ${\textbf{\textit{v}}}_{i}=v_0[\cos\theta_i(t)\hat{\textbf{\textit i}}+\sin\theta_i(t)\hat{\textbf{\textit j}}]$ along the intrinsic motion, being $v_0$ the magnitude of the self-propulsion. The Gaussian white noises ${\boldsymbol{\xi}}_i(t)$ and $\eta_i(t)$ represent the thermal noise, from ordinary Brownian motion, and random torque (due to some inner mechanism of the particles) responsible for changing the internal motion direction, respectively. They have zero means and correlations given by $\langle{\xi_{ia}(t)\xi_{jb}(t')\rangle=2D_t\delta_{ij}\delta_{ab}\delta(t-t^\prime)}$ and $\langle{\eta_i(t)\eta_j(t')\rangle=2D_r\delta_{ij}\delta(t-t^\prime)}$ with intensities $D_t$ and $D_r$, $i$,$j$ = $1,..., N$, and $a$,$b$ = $x$,$y$. The term $\textbf{\textit{F}}_i=\textbf{\textit{F}}_e+\sum_{j\neq  i}\textbf{\textit{F}}_{ij}$ is the net force on the particle $i$, and $\textbf{\textit{F}}_{ij}=\kappa a_{ij}\hat{\textbf{\textit{r}}}_{ij}$, if $a_{ij}>0$ ($\textbf{\textit{F}}_{ij}=0$ otherwise), is the interparticle interaction, with $a_{ij}=\left(\frac{d_i+d_j}{2}-r_{ij}\right)$ being the overlap distance between particles $i$ and $j$ of diameters $d_i$ and $d_j$, respectively. In our model all particles have the same size, then $d_i=d_j=d$. The system is subject to an external force 
 $\textbf{\textit{F}}_e=-\nabla V(\textbf r_{i})$ which acts only in $y$ and is calculated from the harmonic potential      
 \begin{equation}
V({\bf r}_i)=\frac{1}{2}V_0(y_i-y_0)^2, \label{eq4}
 \end{equation}                       
 where $V_0$ is the confinement strength and $y_0$ the minimum of the potential. The particles move in a space of dimensions $L_x\times L_y$, with $L_x=1000$ and $L_y=10$ with boundary periodic conditions in $x$. In $y$ there is no boundary (the potential keeps the particles in a finite space). This choice for boundary in $y$ yields an effective area fraction of $\phi=\frac{N\pi d^2}{4L_xL_y^\prime}$, where $L_y^\prime$ is a function of $V_0$. In fact, as we will see next, the confinement keeps the particles spread out in a more narrow strip, compared to $L_y$, around $y_0$. The values of the fixed parameters used are:  self-speed magnitude $v_0 = 1$ (which sets the time unit),  particle diameter $d = 1$ (which sets the length unit), $\mu=1$, interparticle repulsion stiffness $\kappa=10$, time step $\Delta t=10^{-3}$, external potential minimum, $y_0=L_y/2=5$, and translational noise intensity, $D_t=0$. With this choice for $D_t$, we follow \cite{Fily}, and neglect thermal fluctuations, since we are interested only in the effects of the active degree of freedom. The study of ordinary brownian particles in confinement was reported in \cite{lucena}. Each simulation is repeated 5 times; averages are taken over the number of runs, and time, where appropriate.     	             

\section{Results and discussions}\label{Results} 

In this section, we present our results for MSD curves, along with other data that will help us to explain our observations.

\subsection{Diffusion}                    	     

%----------------------------------------------------------------- FIGURE ------------------------------------------------------------------------------------------------
%\begin{comment}
\begin{figure*}[!htb]
\centering
\subfigure[]{
\includegraphics[scale=0.3]{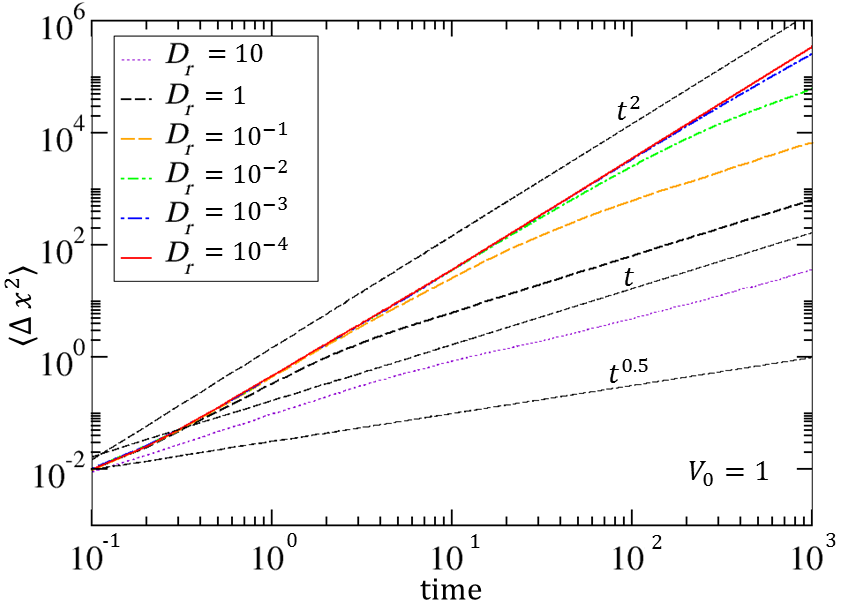}
}
\subfigure[]{
\includegraphics[scale=0.3]{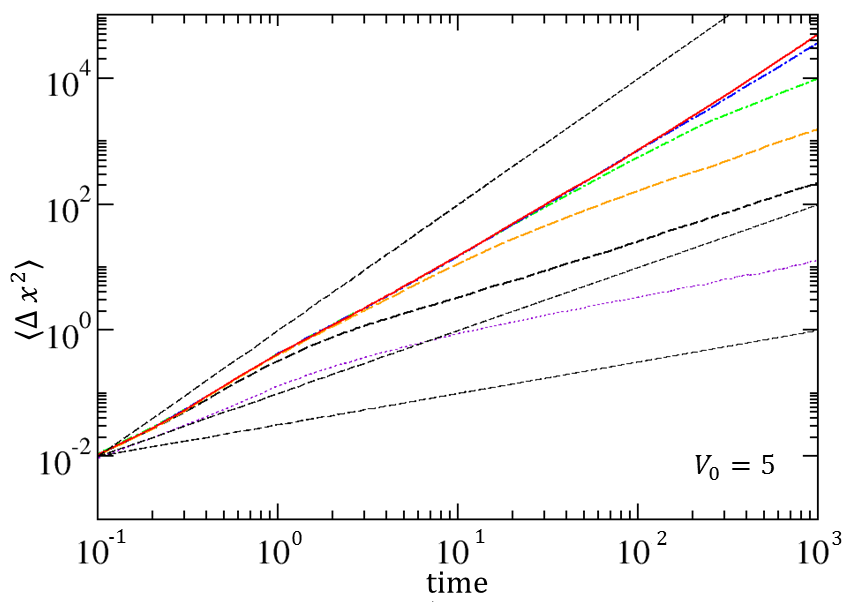}
}
\subfigure[]{
\includegraphics[scale=0.3]{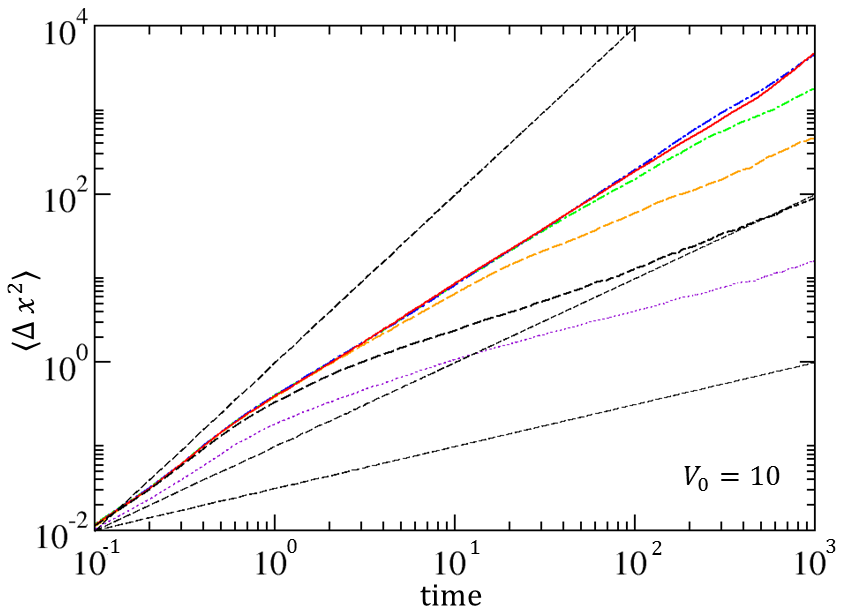}
}
\subfigure[]{
\includegraphics[scale=0.3]{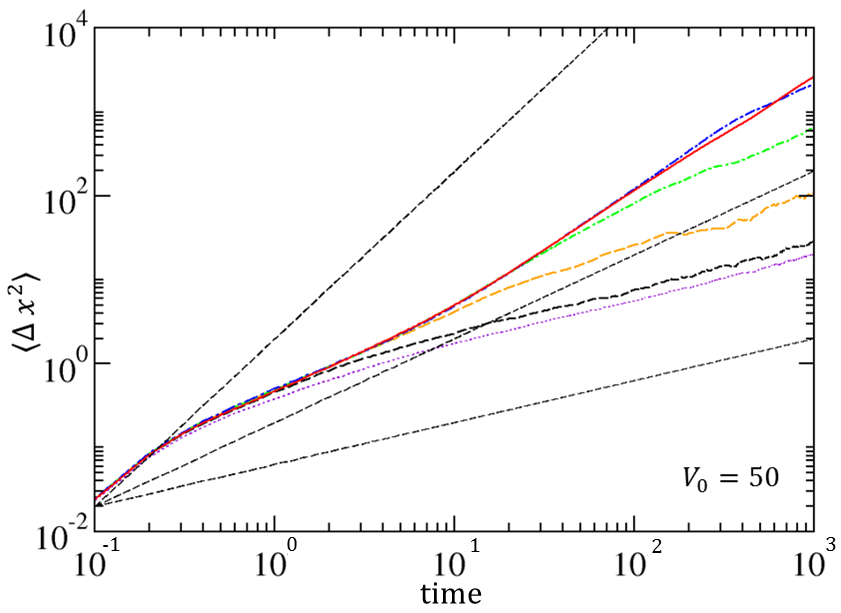}	
}
\caption{(Color online) MSD vs. time as function of $D_r$ and $V_0$. The three straight lines have slopes corresponding to ballistic ($\alpha =2$), normal ($\alpha=1$) and single-file ($\alpha=0.5$) diffusion.\label{fig1}}
\end{figure*}
\begin{figure}[!htb]
	\centering
	\includegraphics[scale=0.32]{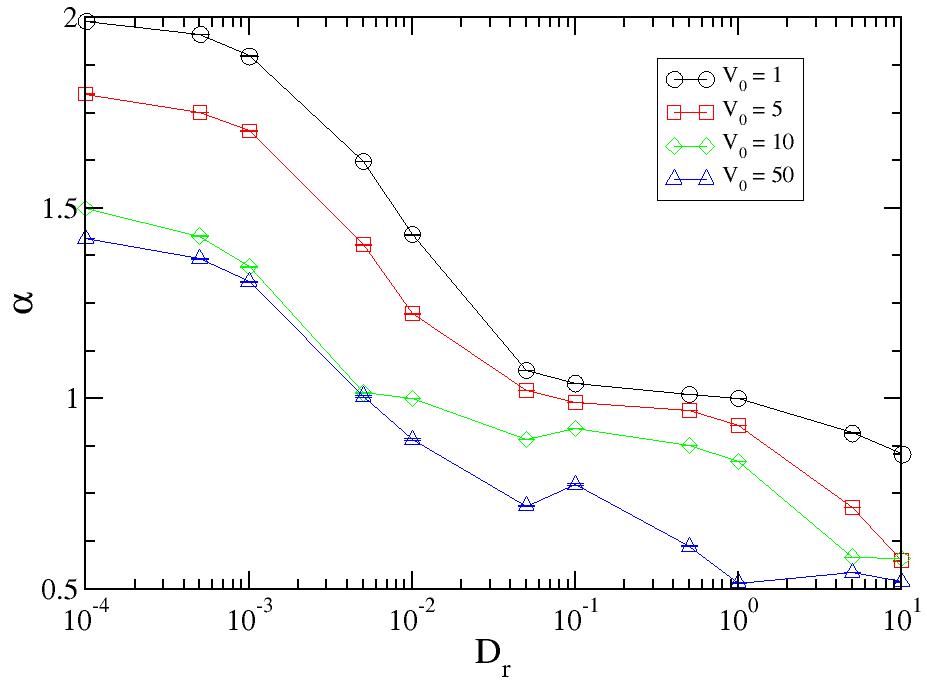} 
	\caption{(Color online) Mean exponent $\alpha$ as a function of $D_r$.\label{fig2}}
\end{figure}

%\begin{figure}[!htb]
%\centering
%\includegraphics[scale=0.4]{expnoise.png} 
%\caption{(Color online) Mean exponent $a$ as a function of $D_r$.\label{fig2}}
%\end{figure}
\hspace*{0.5cm} In fig. \ref{fig1}, we show all MSD curves vs. time for all $V_0$ and $D_r$ values; in fig. \ref{fig2}, we show the mean exponent of each curve, with additional points that were measured for curves that are omitted in fig. \ref{fig1}; these exponents were obtained through power law fits of the curves in fig. \ref{fig1}, in the long time limit, i.e, we only considered for these fits parts of those curves in which we have, at least, time$\geq10$. A feature readily seen is that, for a fixed $V_0$, the exponent $\alpha$ decreases with increasing $D_r$: for $V_0=1$, fig. \ref{fig1}a, we go from ballistic ($\alpha=2$) at $D_r\leq10^{-3}$ to normal diffusion ($\alpha=1$) for $D_r\geq10^{-1}$, even reaching subdiffusion at $D_r=10$; while for higher $V_0$, we begin at superdiffusion in all cases (in particular, $\alpha$ is slightly larger than 1 for $V_0=50$) for $D_r=10^{-4}$ and then the subdiffusive regime is attained ($\alpha=0.5$, the SFD value) at $D_r=10$.\\

The appearance of subdiffusive regimes is simply a consequence of a more restricted space (i.e, denser system) in $y$, for particles to explore, as we will show below. As seen in figs. \ref{fig1} and \ref{fig2}, this compaction effect takes place either as we increase $V_0$ or $D_r$. The effect of the confinement magnitude $V_0$ on the spreading of the particles around $y_0$ is obvious (a larger confinement strength results in a system more compacted around $y_0 = 5$), and an estimate of the width of the region in which the particles are distributed around the potential minimum can be obtained by equating the external force $V_0(y-y_0)$ to the active force $v_0/\mu$ and solving for $y-y_0$ (which is half the size of the region around $y_0$). The calculation yields for $y-y_0$ the following
\begin{equation}
y-y_0=\frac{v_0}{\mu V_0}. \label{eq5}	
\end{equation}    
Hence, a larger $V_0$ implies a more confined (compacted) space for particles to move. The role of $D_r$ on the spatial restriction is not as obvious as the effect of $V_0$; however, we may understand it as a consequence of the fact that active particles with a lower $D_r$ have a larger probability to keep moving along a given direction for a longer time than those with larger $D_r$ (longer persistent motion). Therefore, particles with a lower $D_r$ are able to climb the potential barrier further than those with larger $D_r$, rendering a less compact system. This effect is more readily seen at low $V_0$.
The effects of $V_0$ and $D_r$ just we have described do not explain the unexpected normal and superdiffusive regimes at high confinement; clearly something else should be occurring, which is controlled by these two parameters.\\

\begin{figure}[!htb]
	\centering
\subfigure[]{
\includegraphics[scale=0.32]{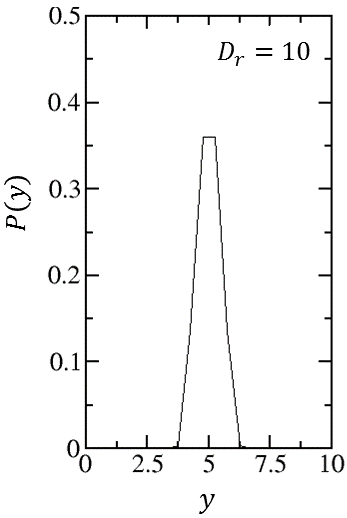}
}
\subfigure[]{
\includegraphics[scale=0.32]{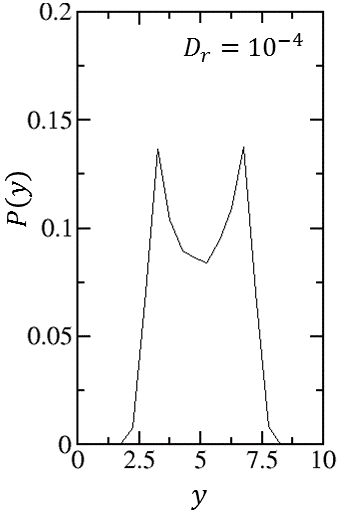}
	}
\subfigure[]{	
\includegraphics[scale=0.32]{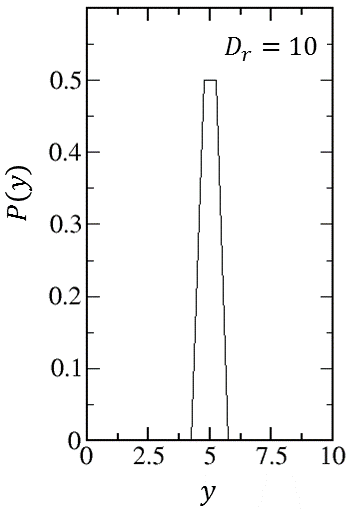}
}
\subfigure[]{	
\includegraphics[scale=0.32]{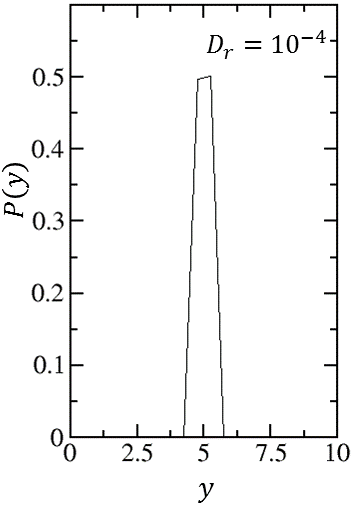}
}
\caption{(Color online) Probability function of particles along $y$ for (a), (b) $V_0=1$ and (c), (d) $V_0=50$.\label{fig3}}
\end{figure}

In order to understand this behavior, we measured the probability distribution of particles along $y$, $P(y)$; we show a few of such curves in fig.\ref{fig3}.%for both $V_0=1$ and $V_0=50$ with $D_r=1$ and $D_r=10^{-4}$.
For fig. \ref{fig3}a ($V_0=1$, $D_r=10$), we see partly the picture described above; particles remain close to $y_0$ (the minimum of the potential), rendering a denser system and a slower diffusion when $D_r$ is large, while they spread out significantly around $y_0$ for low noise and confinement strength (see fig. 3b, $V_0=1$, $D_r=10^{-4}$). Comparing figs. \ref{fig3}a with \ref{fig3}c ($V_0=50$, $D_r=10$) and \ref{fig3}b with \ref{fig3}d ($V_0=50$, $D_r=10^{-4}$), we see the effect of $V_0$ compacting the system; in both cases, the width of $P(y)$ decreases. In general, the effect of either increasing $V_0$ or $D_r$ is to compact the system, as explained above, and is reflected in all $P(y)$ curves that we measured. Finally, the double peak seen in fig. \ref{fig3}b does not appear for $V_0>1$ or $D_r>10^{-2}$. Note that these two peaks are not independent, and we cannot, at this point, interpret this result as evidence of the well-known transition from single file to double file in passive systems \cite{lucena,Piacente10,delfau13}, occurring in this context. We leave this problem to study in a future work.\\

\begin{figure}[!h]
\centering
\subfigure[]{
\includegraphics[scale=0.35]{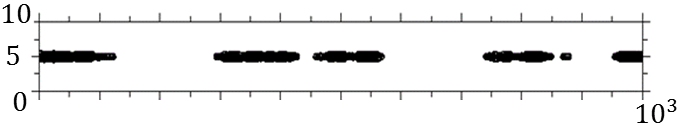}
}
\subfigure[]{
	\includegraphics[scale=0.35]{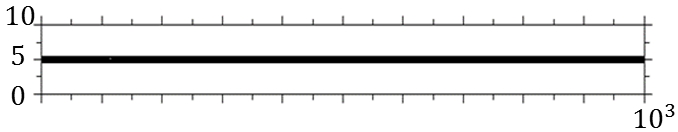}
}
\caption{(Color online) System configuration for $V_0=50$ and (a) $D_r=10^{-4}$ and (b) $D_r=10$.\label{fig4}}
\end{figure}

Comparing figs. \ref{fig3}c and \ref{fig3}d, we do not see any significant change in $P(y)$. In fact, for $V_0=50$ this probability hardly changes with $D_r$. Nevertheless, we have sub (practically the SFD regime) and superdiffusion for $V_0=50,D_r=10$, and $V_0=50,D_r=10^{-4}$, figs.  \ref{fig3}c and \ref{fig3}d, respectively.
In order to understand what is going on in this high confinement regime, we show two snapshots of the steady state of the system in fig. \ref{fig4}, for the parameters in figs. \ref{fig3}c, and \ref{fig3}d.
\\
              
Now, the situation becomes clearer: at high $V_0$, and low $D_r$, there is an aggregation effect taking place, while these structures seem to be absent at high $D_r$. In this high $V_0$, low $D_r$ regime, the particles cannot climb the potential barrier, as previously seen; they, instead, stay around $y_0$, and form clusters, as a consequence of their longer persistent motion. Therefore, we see that super and normal diffusion seen at high $V_0$ is the result of the available space that is left among clusters which allows them to diffuse more swiftly than at the (apparently) SFD picture seen at fig. \ref{fig4}b, (a similar effect was reported for magnetic brownian particles in confinement \cite {lucena14} in which attractive interactions were responsible for the aggregation). In the next section, we study these clusters a little further so that we can to understand their role in the system.   

\subsection{Clustering}
\begin{figure}[!htb]
\centering
\includegraphics[scale=0.3]{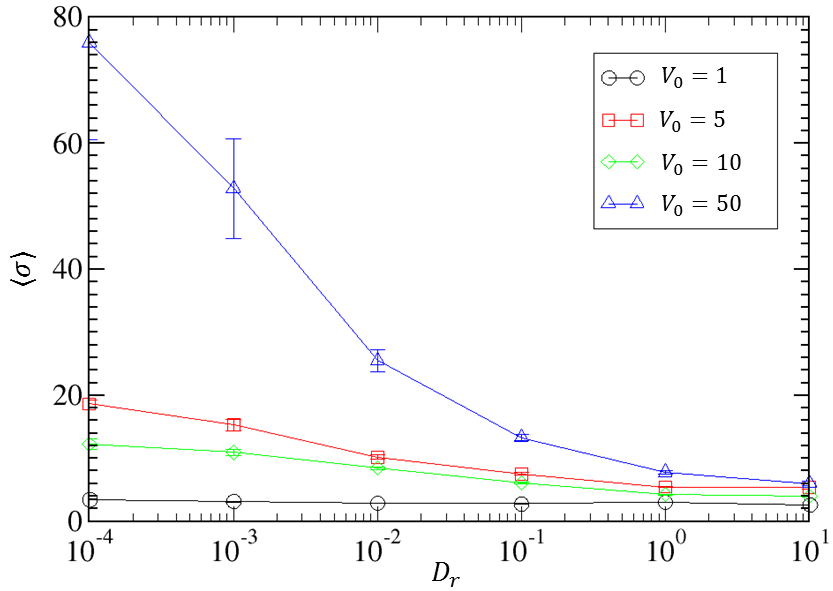}
\caption{(Color online) Average cluster size as function of $D_r$.\label{fig5}}
\end{figure}

Here, we show our results concerning the average cluster size $\langle\sigma\rangle$, fig. \ref{fig5}, as a function of $V_0$ and $D_r$. We considered that two particles form a cluster if their distance is $r\leq1.01$. From fig. \ref{fig5}, we see that for $V_0=1$ there are only small clusters, since $\langle\sigma\rangle$ is about 3 for any noise strength, and it is fairly insensitive to variations in $D_r$. For the other confinement strengths, we see the same qualitative dependence on $D_r$: it grows with decreasing noise intensity. For $V_0=50$ and $D_r=10^{-4}$, we have $\langle\sigma\rangle\approx80$, which means a discernible local structure (with only a handful of clusters at any time, since $N$ is fixed). By itself, these results only imply the existence of these aggregates, and do not mean more available space for diffusion; nevertheless, we see that as we increase $V_0$, a larger diffusion exponent is obtained where we have few, large clusters.\\

\begin{figure}[!htb]
\centering
%\begin{minipage}{\textwidth}
\subfigure[]{
\includegraphics[scale=0.32]{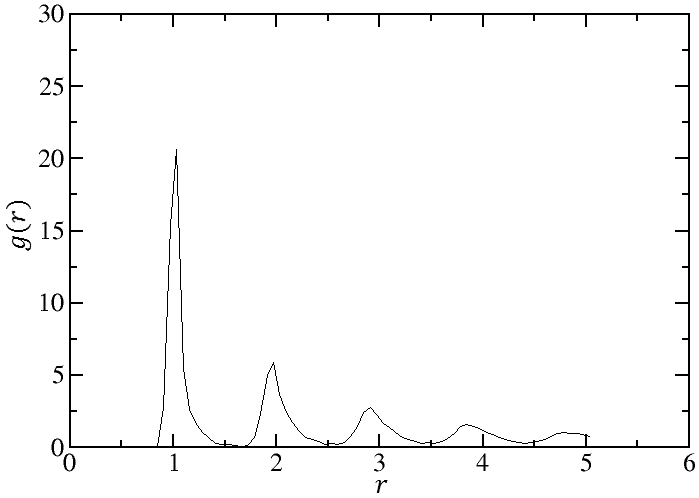}
}
\subfigure[]{
\includegraphics[scale=0.32]{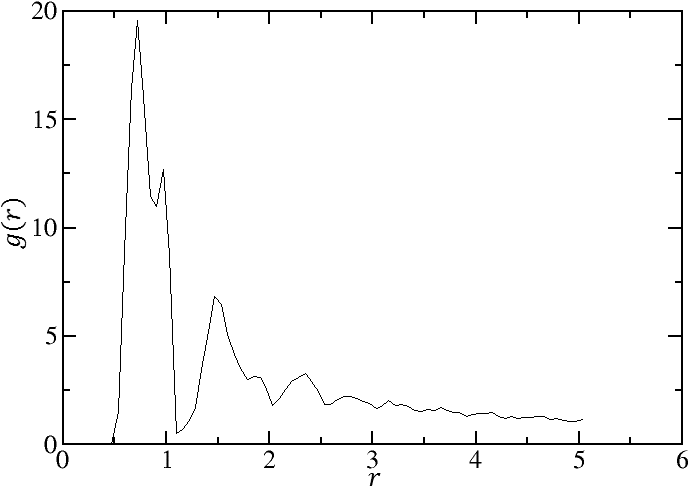}
}
%\end{minipage}{\textwidth}
\caption{(Color online) Radial distribution function for $V_0=50$ and (a) $D_r=10$ and (b) $D_r=10^{-4}$.\label{fig6}}
\end{figure}

In order to understand the change in the structure of the system, we computed the radial distribution function, $g(r)$, displayed in fig. \ref{fig6}. In fig. \ref{fig6}a, we can see a sequence of peaks around $r=1,2,3,...$. This is only possible if particles are arranged in a line, which gives additional support for SFD behavior seen at $V_0=5,10,50$ for $D_r=10$ (in fact, $g(r)$ for these three cases are similar to each other). In fig. \ref{fig6}b, these peaks are absent, and there is no evidence of long range structure; however, we see a high and wide peak around $r\approx0.5$, what means a strong overlap among particles, which it turns out to be the signature, in the structure, of the clusters seen in fig. \ref{fig4}a. Given the size and location of the first peak, we may infer that there should be a significant amount of empty space among the clusters. This is reflected in the structure of the system in that the RDF quickly decreases below its normalization value for $r>2$. It should be stressed that such strong overlaps are a consequence of the weak interparticle repulsion, i.e, small stiffness, $\kappa$. Although we did not investigate whether there is significant, if any, mutual passage (what characterizes the SFD), we may speculate that particles with higher $\kappa$ will overlap less and, therefore, occupy the empty space which is seen in figs. \ref{fig4}a and \ref{fig6}b. Since the empty space among the clusters is what allows the normal and superdiffusion to occur, we may speculate that in hard sphere limit ($\kappa\rightarrow\infty$) we will only observe SFD for strong confinement.        

\section{Conclusions and Outlooks}\label{Conclusions}

We studied in this letter the diffusive properties of an active matter system confined laterally by a harmonic potential. We measured the mean square displacement (MSD) in the $x$-direction for several confinement, $V_0$, and noise, $D_r$, strengths.\\

The diffusion curves showed an inverse dependence between $\alpha$, the diffusion exponent, and $D_r$ for a given $V_0$. We showed that the subdiffusive regimes arise due to compaction of the system when we increase $V_0$ and $D_r$. Our data for the radial distribution function at high $V_0$ gave us evidence that the MSD curve with a diffusion exponent of $\alpha=0.5$ accounts for a linear chain of particles as seen by the secondary peaks at integer distances, yielding a strong evidence for a SFD regime.\\

On the other hand, we observe normal and superdiffusion at all $V_0$, for low $D_r$. For low confinement, particles span a wide stripe around $y_0$, the potential minimum, which renders an effective dilute system. For high $V_0$, we observed that particles form clusters, leaving a significant amount of empty space among them, allowing a faster diffusion. In both cases, these effects occur due to longer persistent motion of the particles. We inferred the second fact from snapshots of the system configuration, along with the data for $g(r)$  which show a high and wide first peak about $r\approx0.5$ at high $V_0$ and low $D_r$. The strong overlap among particles, which is the signature of the clusters, is due to weak mutual repulsion, low $\kappa$. For harder particles, we may speculate that the clusters will disappear, and only subdiffusion (although not necessarily SFD), at high $V_0$, will take place.
% We hope this work contributes for a better understanding on active matter systems and that it can to serve as a path for further works that approach the effect of fluctuations and spatial restrictions in such a systems.                

\section{Acknowledgments} 
We would like to thank prof. W. P. Ferreira for discussing with us about this problem. We also would like thank to CAPES and FAPESPA for financial support.

\end{document}